\documentclass[10pt,letterpaper]{article}
\usepackage[top=0.85in,left=2.75in,footskip=0.75in]{geometry}

\usepackage{amsmath,amssymb}

\usepackage{changepage}

\usepackage[utf8x]{inputenc}

\usepackage{textcomp,marvosym}

\usepackage{cite}

\usepackage{nameref,hyperref}

\usepackage{microtype}
\DisableLigatures[f]{encoding = *, family = * }

\usepackage[table]{xcolor}

\usepackage{array}

\newcolumntype{+}{!{\vrule width 2pt}}

\newlength\savedwidth

\usepackage{setspace} 
\raggedright
\usepackage[aboveskip=1pt,labelfont=bf,labelsep=period,justification=raggedright,singlelinecheck=off]{caption}

\makeatletter
\renewcommand{\@biblabel}[1]{\quad#1.}
\makeatother

\usepackage{lastpage,fancyhdr,graphicx}
\usepackage{epstopdf}
\fancyheadoffset[L]{2.25in}
\fancyfootoffset[L]{2.25in}
\begin{document}
\vspace*{0.2in}

\begin{flushleft}
{\Large
\textbf\newline{Randomly connected networks generate emergent selectivity and predict decoding properties of large populations of neurons} % Please use "sentence case" for title and headings (capitalize only the first word in a title (or heading), the first word in a subtitle (or subheading), and any proper nouns).

}

% \newline
% \\
Audrey Sederberg\textsuperscript{1,2},
Ilya Nemenman\textsuperscript{1,2,3}
% Name3 Surname\textsuperscript{2,3\textcurrency},
% Name4 Surname\textsuperscript{2},
% Name5 Surname\textsuperscript{2\ddag},
% Name6 Surname\textsuperscript{2\ddag},
% Name7 Surname\textsuperscript{1,2,3*},
% with the Lorem Ipsum Consortium\textsuperscript{\textpilcrow}
\\
\bigskip
\textbf{1} Department of Physics, Emory University, Atlanta, GA, USA 
\\
\textbf{2} Initiative in Theory and Modeling of Living Systems, Emory University, Atlanta, GA, USA
\\
\textbf{3} Department of Biology, Emory University, Atlanta, GA, USA
% \\

% \\
% \textbf{3} Affiliation Dept/Program/Center, Institution Name, City, State, Country
% \\
% \bigskip

% Insert additional author notes using the symbols described below. Insert symbol callouts after author names as necessary.
% 
% % Remove or comment out the author notes below if they aren't used.
% %
% % Primary Equal Contribution Note
% \Yinyang These authors contributed equally to this work.

% % Additional Equal Contribution Note
% % Also use this double-dagger symbol for special authorship notes, such as senior authorship.
% \ddag These authors also contributed equally to this work.

% % Current address notes
% \textcurrency Current Address: Dept/Program/Center, Institution Name, City, State, Country % change symbol to "\textcurrency a" if more than one current address note
% % \textcurrency b Insert second current address 
% % \textcurrency c Insert third current address

% % Deceased author note
% \dag Deceased

% % Group/Consortium Author Note
% \textpilcrow Membership list can be found in the Acknowledgments section.

% Use the asterisk to denote corresponding authorship and provide email address in note below.
* audrey.j.sederberg@emory.edu

\end{flushleft}
\section*{Abstract}

Advances in neural recording methods enable sampling from populations of thousands of neurons during the performance of behavioral tasks, raising the question of how recorded activity relates to the theoretical models of computations underlying performance. In the context of decision making in rodents, patterns of functional connectivity between choice-selective cortical neurons, as well as broadly distributed choice information in both excitatory and inhibitory populations, were recently reported \cite{Najafi354340}. The straightforward interpretation of these data suggests a mechanism relying on specific patterns of anatomical connectivity to achieve selective pools of inhibitory as well as excitatory neurons. We investigate an alternative mechanism for the emergence of these experimental observations using a computational approach. We find that a randomly connected network of excitatory and inhibitory neurons generates single-cell selectivity, patterns of pairwise correlations, and indistinguishable excitatory and inhibitory readout weight distributions, as observed in recorded neural populations. Further, we make the readily verifiable experimental predictions that, for this type of evidence accumulation task, there are no anatomically defined sub-populations of neurons representing choice, and that choice preference of a particular neuron changes with the details of the task. This work suggests that distributed stimulus selectivity and patterns of functional organization in population codes could be emergent properties of randomly connected networks.

% Please keep the Author Summary between 150 and 200 words
% Use first person. PLOS ONE authors please skip this step. 
% Author Summary not valid for PLOS ONE submissions.   
\section*{Author summary}
What can we learn about neural circuit organization and function from recordings of large populations of neurons? For example, in population recordings in the posterior parietal cortex of mice performing an evidence integration task, particular patterns of selectivity and correlations between cells were observed. One hypothesis for an underlying mechanism generating these patterns is that they follow from intricate rules of connectivity between specific neurons, but this raises the question of how such intricate patterns arise during learning or development. An alternative hypothesis, which we explore here, is that such patterns emerge from networks with broad spectra of eigenvalues, which is a generic property of certain random networks. We find that a random network model matches many features of experimental recordings, from single cells to populations. We suggest that such emergent selectivity could be an important principle in brain areas, in which a broad distribution of selectivity is observed. 

%%%%%%%%% uncomment
% \linenumbers

% Use "Eq" instead of "Equation" for equation citations.
\section*{Introduction}
With the deluge of neural recordings made possible by modern recording methods \cite{Stringer2019}, theoretical neuroscience must address the challenge of relating complex activity patterns to the algorithms thought to underlie brain function \cite{Najafi2018, Brody2016}. For example, evidence accumulation tasks explore how the brain makes decisions based on the temporal integration of incoming sensory information. One class of models for performing this discrimination is that of attractor networks. In an attractor network model of decision making, pools of neurons fire selectively for a particular choice. A transient activation is prolonged through slow recurrent excitation within the pool while inhibiting other pools of neurons that are selective in their firing for other choices through non-specific inhibition \cite{Wang2002}. The results of recent experiments in decision-related areas of rodent parietal cortex have called this model into question \cite{Raposo2014, Licata2017, Najafi354340}. These experiments showed that, contrary to the predictions of the original models, inhibitory cells are also selective for choice, suggesting an alternative mechanism, in which pools of inhibitory neurons selectively inhibit neurons representing evidence for the opposite choice \cite{Aksay2007, Najafi354340}. However, it is unclear how the specific pattern of connectivity would be generated. In this paper, we explore an alternative mechanism for these experimental observations. 

More specifically, in this task \cite{Najafi354340}, a rodent is presented with an irregular train of either visual or auditory impulses and must determine whether the average frequency with which those pulses arrived is above or below some internally remembered threshold. Recordings of population activity in the posterior parietal cortex (PPC) during the task revealed weak choice selectivity in single cells, with a fraction of individual cells showing significant selectivity for one of the two choices.  A linear classifier operating on the activity across the population decoded the choice with high accuracy. Both excitatory and inhibitory neurons were selective for choice, and noise correlations between pairs of neurons reflected whether stimulus selectivity was shared or opposing. A straightforward mechanism for these observations is that some specific pattern of connectivity exists in the cortical network that separates inhibitory cells into ``pools'' that are selective for specific choices.  

We explore an alternative mechanism for the emergence of these observations. We hypothesize that a randomly connected network can produce patterns of activity that are sufficiently distinct to differentiate input frequency. The eigenvalue spectrum of a randomly drawn connectivity matrix often has a tail of eigenvalues with large real parts \cite{Mehta2004}, and this remains true even when networks follow Dale's principle of separate excitatory and inhibitory neurons \cite{Rajan2006, Aljadeff2015}. We reason that input modes overlapping with fast-growing modes are amplified by network dynamics, and nonlinear input-output function at the level of single cells could result in an the activation pattern that depends on the temporal frequency of the input. Thus, the network would produce different patterns of activation as a function of the input frequency, resulting in emergent selectivity for frequency at the level of the population. We are primarily interested in whether such a generic random network of firing rate units could generate the observed patters of selectivity, functional connectivity, and cell-type-specific readout weight distributions simply through randomly arising heterogeneity in synaptic inputs at the level of single neurons. A thorough theoretical derivation of selectivity and readout weight distributions would be of interest but rather complex, so instead we explore this question computationally. 

Our simulation results support this hypothesis. Our main findings are (i) that heterogeneity in connectivity generates differences in inputs to single cells that are dependent on stimulus frequency and (ii) that these differences are sufficient to distinguish between low- and high-frequency inputs. Our model reproduces experimental findings, including the distribution of single-neuron selectivity, patterns of pairwise noise correlations, the performance of a classifier, and the distribution of readout weights. Our theory makes the verifiable experimental predictions that, if the mechanism is through emergence rather than specific connectivity, then (1) there is no anatomical basis for sub-populations tuned to a particular choice and (2) when task parameters, such as input frequency are changed, neural selectivity also changes. These results suggest a mechanism for how cortical networks could exhibit functional organization without specific patterns of cortical connectivity.

\section*{Results}
\subsection*{Simulation of a temporal evidence accumulation task}
We first review key experimental results for a particular version of an evidence accumulation task that can be performed by rodents  \cite{Raposo2014, Licata2017, Najafi354340}. In these tasks, rodents are trained to discriminate low- and high-frequency inputs (pulses, either visual or auditory) and to report their choice. The exact timing of the inputs is random, so the rodent is not reporting the inter-pulse interval but deciding based on the stimulus history over a short period, typically one second. They do this accurately. Recordings of neurons in PPC during the task show a distribution of selectivity for choice, meaning that the reported outcome of the evidence accumulation is predictable from the activity of some fraction of the neurons. A classifier trained to determine choice from the recorded population activity performed accurately. Interestingly, it performed equally well for excitatory or inhibitory sub-populations, and an analysis of the classifier weights showed that excitatory and inhibitory cell activities were weighted similarly \cite{Najafi354340}. 

\begin{figure}[htbp]
\includegraphics[width=\textwidth, trim ={0cm 11.5cm 0cm 0cm},clip=true]{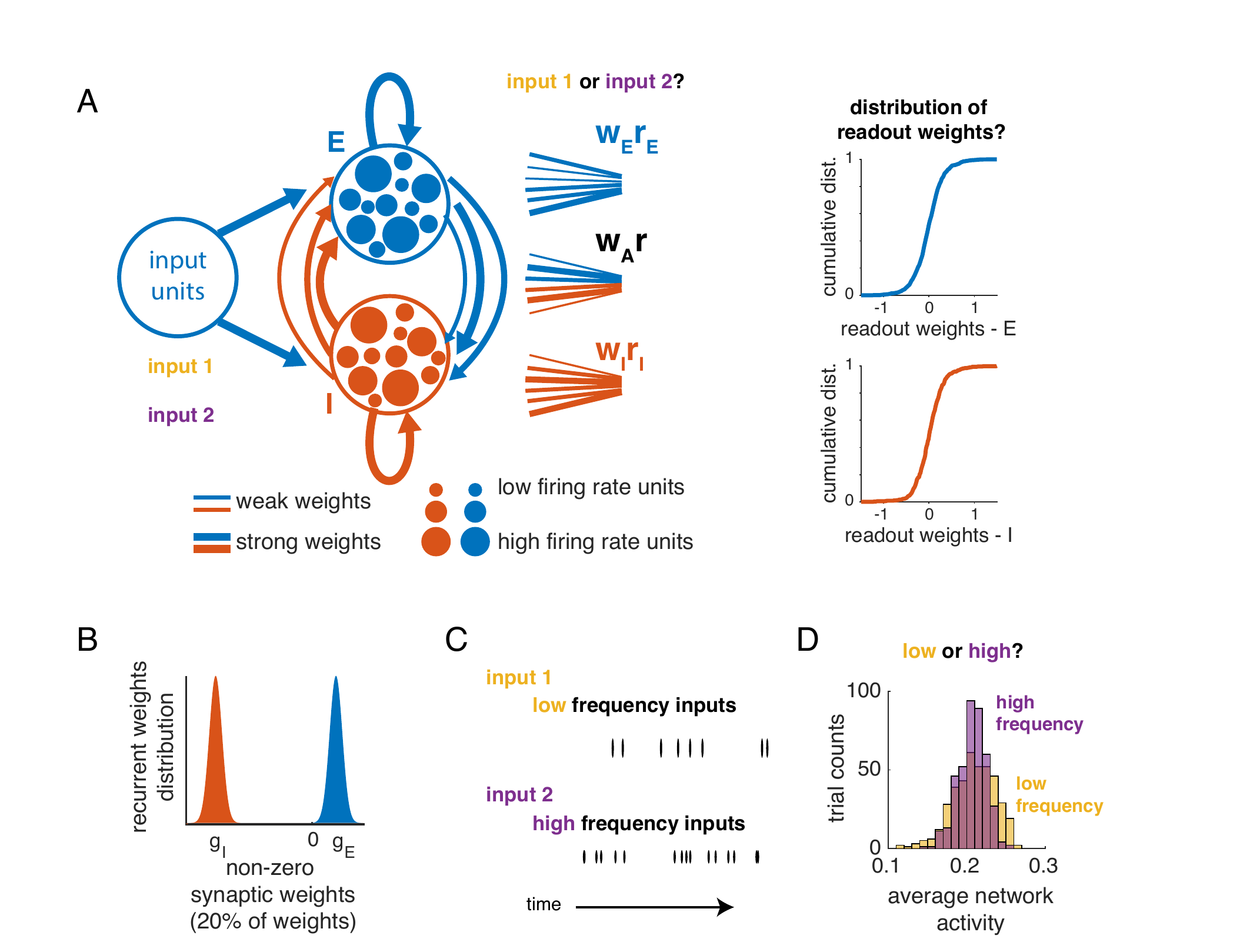}
\caption{ {\bf Overview: Simulation of an input frequency discrimination task in a recurrent network of excitatory and inhibitory neurons with random connectivity.}
A: Recurrent E-I network of randomly connected neurons. Synaptic weights are heterogeneous (line thickness), and connectivity in the network is random. Inputs to the network have the same pattern of spatial activation but different temporal features. Readouts of network activity (from all neurons, or restricted to excitatory or inhibitory neurons) discriminate between stimulus categories. We then analyze the distributions of readout weights. 
B: Distributions of non-zero inhibitory (shaded red) and excitatory (shaded blue) synaptic weights. 20\% of the weights are non-zero. C: Inputs to the network are pulses arriving at random times with low (top row, yellow) or high (bottom row, red) average frequency. 
D: Histogram of population average network activity over trials for low- and high-frequency stimuli. Average network activity is not informative of input categories.  }
\label{fig:f1_setup}
\end{figure}

\begin{figure}
\includegraphics[width=0.75\textwidth, trim ={1cm 1cm 1cm 0cm},clip=true]{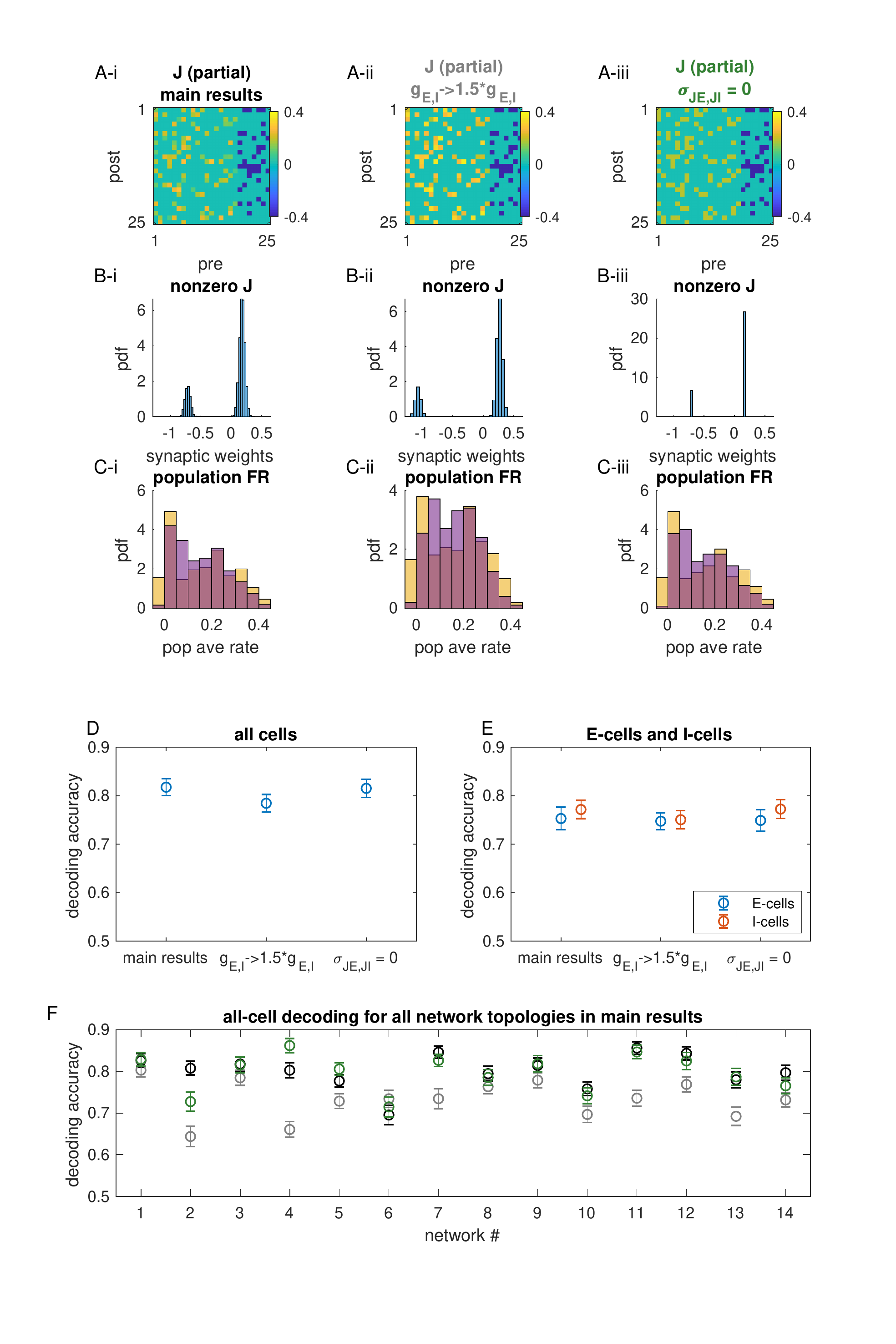}
\label{S2_Fig}

{\bf S2 Figure. Additional parameter combinations.} Fixing network topology (i.e., which elements of J are non-zero), we simulated three networks: with original weights (i), with all synaptic weights scaled by a factor of $1.5$, (ii) and with homogeneous excitatory and homogeneous inhibitory synaptic weights (iii). A: Image of network connectivity for 25 (of 500 total) neurons showing that the topology was kept the same for each simulation. B: Distribution of non-zero excitatory and inhibitory weights in the network. Note that there are approximately four times as many excitatory weights, but they are on average a quarter of the strength of inhibitory weights. C: Histogram of stimulus 1 (yellow) and stimulus 2 (purple) population firing rates for each parameter scaling. Firing rates are matched for each simulation individually; these are operating over a similar population firing rate range. D: Decoding accuracy of the full population in each network. E: Decoding accuracy of excitatory and inhibitory cells in each network. D and E show that decoding accuracy persists after a drastic parameter change, for this network topology. F: Decoding performance for all simulated networks. 
\end{figure}

Here we study a randomly connected rate-based network of $N_E$ excitatory cells and $N_I$ inhibitory cells that performed the evidence accumulation task described above (Fig.~\ref{fig:f1_setup}A). While similar in some aspects, this network is not set up as required for reservoir computing \cite{Jaeger78, Maass2002}, in that our network operates in a regime in which spontaneous activity is low in the absence of inputs. The firing rate of individual units is the sum of any external input, representing the stimulus, and synaptic inputs from the rest of the network, passed through a saturating non-linearity so that activity is between 0 and 1.  Connectivity in the network is sparse, with a 20\% probability of connection between two cells. Excitatory and inhibitory synaptic weights are drawn from truncated normal distributions ($\sim N(g_{E,I}, \sigma^2)$) (Fig.~\ref{fig:f1_setup}B, Methods), and excitatory and inhibitory synaptic weights to neurons are balanced on average, such that $N_E g_E - N_I g_I = 0$. These are set to generate a broad spectrum of eigenvalues in the connectivity matrix \cite{Rajan2006}. We fix $g_E$, $g_I$ and $\sigma$, and we focus on a single combination of parameters, but the essential results are not dependent on making these particular choices (see \nameref{S2_Fig} for additional parameter choices). The stimuli consist of pulsed inputs arriving at random times (Fig.~\ref{fig:f1_setup}C, see Methods) and with average frequency $f$. There is no spatial component of the task: inputs stimulate the same subset of excitatory input neurons in all trials. From the simulated activity of the network, we decode the frequency of the input, using either the full population or only excitatory cells or inhibitory cells. 

This network was conceived as a model of the posterior parietal cortex (PPC) of rodents, which does not receive direct sensory inputs but rather receives inputs that have passed through upstream networks \cite{Whitlock2017}. Moreover, the average population firing rates in PPC during such a task are not directly related to stimulus frequency \cite{Raposo2014}. To account for this effect in our simulation, we scaled the amplitude of inputs such that the network firing rate is, on average, equal for each frequency (Fig.~\ref{fig:f1_setup}D and \nameref{S1_Fig}). From a computational standpoint, this choice makes the task of decoding from the network activity more difficult, as the average firing rate does not encode stimulus identity. 

We present the simulation results by following the experimental observations presented by Najafi and colleagues \cite{Najafi354340}. First, we analyze simulated choice selectivity at the level of single cells and pairs of cells. Given that we know the network connectivity, we analyze both noise correlations and the underlying connectivity pattern. Next, using the simulated population activity, we decode choice from the simulated population activity, and we compare the distribution of weights from the readouts of simulated activity to the distribution acquired for experimentally recorded activity. Finally, we simulate new conditions, in which we change the frequencies being discriminated, and from this simulation, we predict how this changes the selectivity for choice, both in single neurons and across the population.

\subsection*{Emergent selectivity for stimulus category in single cells}
\begin{figure}[htbp]
\includegraphics[width=\textwidth, trim ={0cm 0cm 1cm 0cm},clip=true]{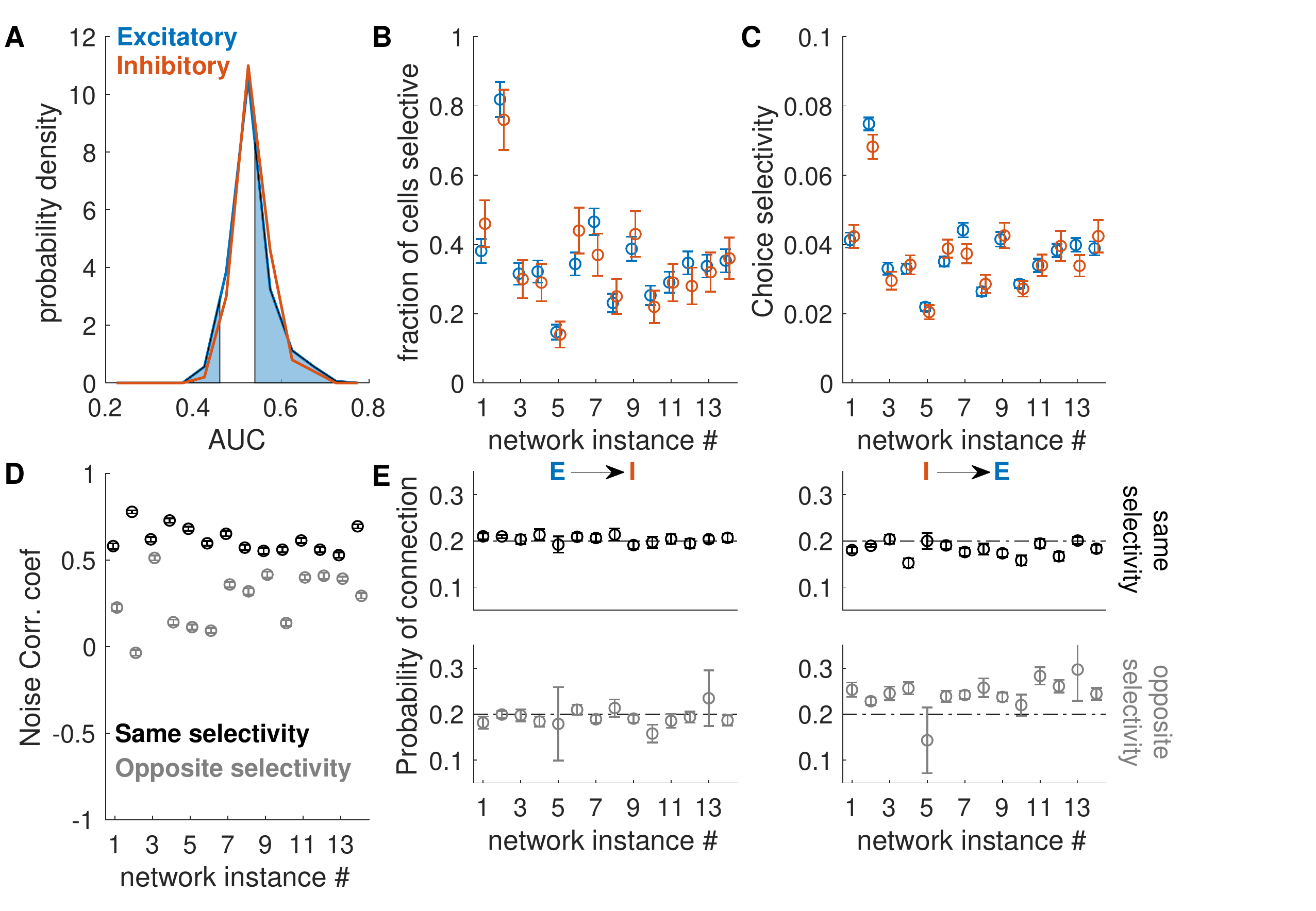}
\caption{ {\bf Weak selectivity in both excitatory and inhibitory cells.}
A. Distribution of AUC values for a single simulation, for excitatory (blue) and inhibitory (red) cells. Shaded regions are those AUC values exceeding significance bounds generated by shuffling trial labels, shown for excitatory cells only. Bounds for inhibitory cells are similar.
B. The fraction of cells selective is the fraction of cells in the excitatory or inhibitory populations whose AUC exceeded the significance bounds. The fraction was variable across network realizations, but within each network, it was similar for excitatory and inhibitory cells. Error bars are derived from counting statistics. 
C. Average choice selectivity (defined as in \cite{Najafi354340} as the average of $|AUC - 0.5|$). Averages are over all cells (selective and non-selective) and error bars are SEM.
D.  Noise correlations (average value) among cells that share the same selectivity is higher than among cells that have opposite selectivity. Error bars are SEM. Noise correlations are substantial ($>0.5$) because there is no internal noise in the network (all noise is input-driven, and shared across cells).
E. Probability of connection for four combinations of pairs of cell types and selectivity: excitatory to inhibitory, with the same selectivity; excitatory to inhibitory, with opposite selectivity; and inhibitory to excitatory with same and opposite selectivity. Error bars are derived from counting statistics.}
\label{fig:f2_single}
\end{figure}

We first examine choice selectivity in single cells from the network simulations. For this analysis, choice was defined to be the correct stimulus label (i.e., low vs.\ high frequency of input pulses) on each trial, which is equivalent to analyzing the correct trials only in a behavioral experiment. Single-neuron activity was averaged over time, yielding a single number per trial for each cell. An ideal observer analysis was used to discriminate between low- and high-frequency inputs (see Methods). Choice selectivity \cite{Najafi354340} was defined as the area under the receiver-operator curve (AUC, \cite{Green1966}), which will be less (greater) than 0.5 when the cell is selective for the low-frequency (high-frequency) stimulus. For each network realization, this generated a distribution of AUC values across all cells (Fig.~\ref{fig:f2_single}A). To assess the significance of a single AUC calculation, we computed the 95\% confidence interval of AUC values obtained with shuffled trial labels. Values exceeding these bounds are significant. Approximately 30\% of excitatory and 30\% of inhibitory cells in the example network in (Fig.~\ref{fig:f2_single}A) had significant selectivity, based on the AUC analysis (Fig.~\ref{fig:f2_single}B, network instance 1). Across different simulated networks, we observed proportions from $0.14$ to $0.82$ of single cells that had significant AUC values, and there was no difference between excitatory and inhibitory cells (Average fraction selective, $N = 14$ networks: $0.36 \pm 0.15$ (excitatory cells) and $0.35 \pm 0.15$ (inhibitory cells)). We also computed the average choice selectivity, defined as $|AUC-0.5|$ for each network (Fig.~\ref{fig:f2_single}C). Across networks, the average choice selectivity was $0.04 \pm 0.01$ in both excitatory and inhibitory cells ($N = 14$ networks, range 0.02 to 0.07), compared to the reported values 0.05 to 0.08 \cite{Najafi354340}. To summarize, in the model, single cells were weakly selective for choice, and excitatory and inhibitory cells exhibited similar levels of selectivity. Thus, we have set up the network such that the average population response is not strongly selective for choice, and in this regime, single cells across the population have selectivity values that are comparable to experimentally recorded values.

\subsection*{Noise correlations reflect relative selectivity of pairs of cells}
We next asked whether this simple, unstructured network also explained pairwise relationships observed in the experimental recordings. Specifically, pairs of cells in PPC that shared the same selectivity had higher noise correlations \cite{Najafi354340} than pairs that had opposite selectivity. To compare this to our simulation results, we computed noise correlation as the cross-correlation between neurons of stimulus-specific responses, averaged across stimuli. Because the only source of noise in the simulation, the variable timing of inputs, is a shared input to all neurons, we expect noise correlation in the model to be higher in the model than in the data, but cells with the same selectivity are expected to have higher noise correlation. Restricting analysis to the cells that exceeded the significance criterion for AUC values, we categorized cells by selectivity and compared average noise correlation between pairs of cells with same and opposite preference. As observed experimentally, noise correlation in the simulation was higher between pairs of cells with the same selectivity than between cells with opposite selectivity (Fig.~\ref{fig:f2_single}D). Thus, organization of functional connectivity in the network emerged without setting up distinct clusters of connections in the network. 

We further examined the patterns of connectivity between these sets of cells (Fig.~\ref{fig:f2_single}E). The overall probability of a synapse was set to 20\% for all simulations. There was a nearly identical probability of connection from an excitatory cell to an inhibitory cell with the same selectivity, $20\% \pm 1\%$ (SD), as with opposite selectivity, $19\% \pm 3\%$ (SD). The inhibitory to excitatory connection between cells with the same selectivity was similar as well, $18\% \pm 2\%$ (SD). Among cells with opposite selectivity, the probability of connection from an inhibitory to an excitatory cell was $24\% \pm 4\%$ (SD). Even this small amount of excess connectivity between inhibitory and excitatory cells of opposite selectivity was sufficient to reproduce the experimentally observed trends in functional connectivity and stimulus selectivity patterns. We emphasize that this bias in connectivity was not put in by hand, but rather was uncovered by the dynamics that it shaped. 

\subsection*{Decoding from population activity}
\begin{figure}[htbp]
\includegraphics[width=\textwidth, trim ={0cm 0cm 0cm 0cm},clip=true]{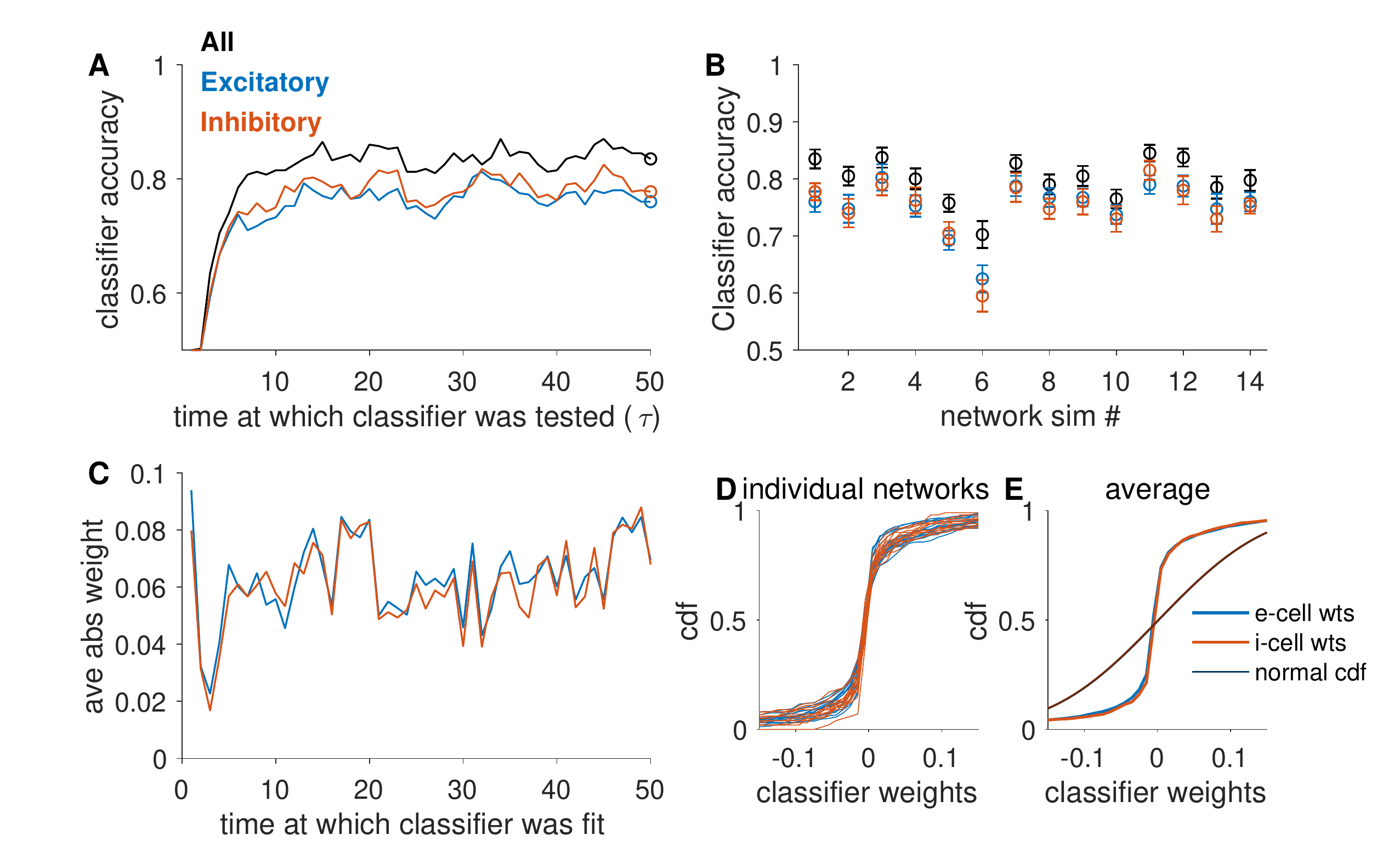}
\caption{ {\bf Classifiers decode stimulus identity from recurrent network activity.}
A: Accuracy of a classifier over the course of the stimulus presentation. For each set of cells (all, black; 100 e-cells, blue; 100 i-cells, red), we fit the classifier at a single point in time (circle) and classified activity over the trial.
B: Accuracy of classifier across network instances (same order as Figure 2). Error bars are $\pm 1$SE estimated by cross-validation (see Methods). 
C - E: Characterization of classifier weights.
C: Weights for classifiers fit on the activity of all cells at each point in time in the stimulus presentation window. Weights on excitatory units are blue and on inhibitory units are red.
D: Cumulative distribution of weights in each of the networks, for excitatory (blue) and inhibitory (red) cells.
E: Average of CDFs shown in D. Distributions are overlapping for excitatory and inhibitory cells. For reference, we also plot the CDF of a normal distribution with standard deviation matched to that of weight distributions (black).}
\label{fig:f3_decode}
\end{figure}

To determine how accurately the simulated network represented choice, we trained linear classifiers to discriminate between low and high stimulus frequency. We fit a classifier using activity near the end of the stimulus period (circles, Fig.~\ref{fig:f3_decode}A) and tested the classifier over the full stimulus period on a set of reserved trials (Methods). In this simulation, classifier accuracy reached maximal performance within $5\tau$ (time constant of the network) and decoded with $83\% \pm 2\%$ accuracy over the last half of the stimulus period. A classifier fit using only the activity of the inhibitory cells  performed with $76\% \pm 2\%$ accuracy, and a randomly drawn subset of excitatory cells equal in number to the number of inhibitory cells decoded with $78\% \pm 2\%$ accuracy. Across all instances of the network, the accuracy of decoders was comparable for inhibitory cells and excitatory ones (Fig.~\ref{fig:f3_decode}B). The range of performance we observed across randomly drawn networks ($69\%\pm 2\%$ to $86\% \pm 2\%$) was highly consistent with the population decoding accuracy observed in experiments (about $70\%$ to $85\%$, \cite{Najafi354340}). 

For the classifier built from the full population activity, we inspected the readout weights (Methods). Weights onto inhibitory neurons and excitatory neurons were not significantly different (Fig.~\ref{fig:f3_decode}C). In all simulated networks, the distributions of weights for excitatory and inhibitory cells were overlapping (Fig.~\ref{fig:f3_decode}D-E). The weight distributions are not normal (for all network, Lillifors test, $p < 0.001$). Thus, as was reported experimentally, we find that both excitatory and inhibitory cells contribute to stimulus decoding and that readout weights are not significantly different between the two. 

\subsection*{Selectivity in the network under different task conditions} 
\begin{figure}[ht]
\includegraphics[width=\textwidth, trim ={1cm 0cm 1cm 0cm},clip=true]{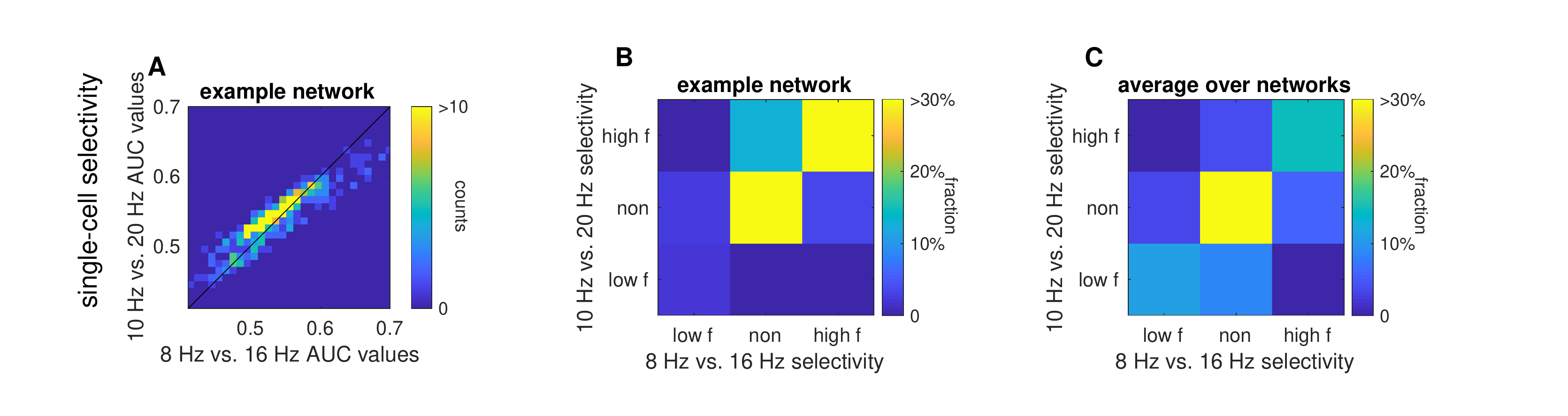}
\caption{ {\bf Changes in selectivity and decoder weights in model when input frequencies are changed.}
A: Density showing AUC (Fig \ref{fig:f2_single}A) for the 8~Hz vs.\ 16~Hz task against the 10~Hz vs.\  20~Hz task, in the same network. Equality line is shown for reference. 
B: Cross-tabulation of selectivity, based on shuffle criterion. Some selective cells become non-selective, and non-selective cells become selective. 
C: Average cross-tabulation of selectivity across all networks. 
}
\label{fig:f4_freqs}
\end{figure}

Finally, we performed a new simulation in which the same network discriminated between different input frequencies. Originally, we set the average frequency of inputs to 8~Hz and 16~Hz. We now ask selectivity changes when the average frequency of inputs changes to 10~Hz and 20~Hz. In all cases, we used the normalization strategy for input amplitudes as before. We then compared the single-cell AUC values for single cells on this new discrimination task to those on the original (8~Hz vs.~16~Hz) task. Fig.~\ref{fig:f4_freqs}A shows the shifts in the selectivity of single cells for an example network. For this simulated network, some cells that had low selectivity on the original task increased their selectivity on the new task, while a subset of selective cells lost selectivity on the new task. We calculated statistical significance for selectivity as in Fig.~\ref{fig:f2_single}. For this network, more cells were selective when the task was to distinguish 10~Hz from 20~Hz than 8~Hz vs.\ 16~Hz (Fig.~\ref{fig:f4_freqs}B), shown by the weight in the off-diagonal entries of the cross-tabulation of selectivity for the original and new task (10~Hz vs.\ 20~Hz). Across different realizations of the network, this was not a strong trend: a fraction of cells either gained or lost selectivity as the task parameters changed (Fig.~\ref{fig:f4_freqs}C). The fraction of all cells that changed selectivity (in either direction) varied across networks but was always greater than zero, averaging 22\% of cells (+/- 9\%, range 8\% to 42\%). To summarize, we predict that the set of selective cells depends on the temporal features of the input and will change if the task changes. 

\section*{Discussion}

We presented a set of simulation results that account for several key features of population recordings in PPC during an evidence accumulation task. This simulated network consists of excitatory and inhibitory neurons with random connectivity. It receives as inputs pulsed sensory signals, which have been filtered by sensory areas. The pulse times are random, with either low or high average rate in time. From the network patterns of activity, we measure single-cell selectivity for input rate, patterns of noise correlations between same and opposite selectivity cells, and readout performance and readout weight distributions. We find that these measures are highly consistent with those observed experimentally. Importantly, our simulations do not include specific patterns of connectivity between excitatory and inhibitory neurons; any biases that emerge are the result of dynamics shaped by random heterogeneity in connectivity patterns. We suspect this is a fairly generic effect of a network with a broad spectrum of eigenvalues, such as the random networks that we studied, but the theoretical connection between the spectra of a connectivity matrix in a nonlinear network and the emergence of selectivity and distribution of readout weights among excitatory and inhibitory populations remains to be explored. 

Experimental work has shown that PPC is integral to performing a sensory evidence integration task \cite{Licata2017, Raposo2014}. Further, the examination of the representation of choice across PPC showed that both excitatory and inhibitory cells have choice selectivity, that cells with the same selectivity had higher noise correlations, and that decoders trained on the population activity patterns to read out choice had both positive and negative weights for both excitatory and inhibitory cells \cite{Najafi354340}. Based on these experimental observations, the authors suggested a model in which multiple pools of excitation and inhibition take as input some variable representing choice and, through specific patterns of connectivity, represent choice across the population, with weak individual selectivity, selective patterns of noise correlations, and zero-mean distributions of decoder weights. We showed that an unstructured network produces all of these effects as well. 

Additionally, previous results have argued that, while the posterior parietal cortex is involved in performing a visual sensory evidence integration task \cite{Licata2017, Raposo2014} in rodents, PPC does not itself integrate evidence \cite{Brody2016, Licata2017}. One possibility, consistent with our simulations, is that the population response is subtly distinct across input frequencies, and these distinctions are learned through a reinforcement mechanism in some other area, which then feeds back to PPC to enhance the distinction between sensory inputs. This feedback mechanism may interact with other biases, such as motivation and trial history \cite{Busse2011, Morcos2016}, which modulate choice. Such a feedback loop could further enhance apparent selectivity in connectivity in PPC because it emphasizes dynamics that were shaped by heterogeneity in connectivity.

Finally, we found that stimulus information could be decoded very early in the stimulus window. The technical reason for this is that due to the input scaling, the first pulse of the input series carried stimulus information. Scaling was used to make the decoding task more challenging and to reproduce the experimental observation that PPC responses to low-frequency inputs are not lower than the PPC responses to high-frequency inputs. We did not implement more realistic adaptation dynamics, which could be the mechanism underlying such scaling. One would expect that such adaptation mechanisms (e.g., short-term depression or facilitation) could be transformed into a population code in a spiking network, readily decoded downstream \cite{Buonomano1995}, and it is an interesting question for future study whether spiking networks such as these replicate the experimentally observable quantities that we focused on here. Adding this feature to the model would  slow the ramp-up in decoding accuracy, but would also require more choices about adaptation rates and tuning. Moreover, such elaboration of the model is superfluous to answering the question of whether a heterogeneous network of excitatory and inhibitory units can distinguish between inputs with different temporal frequencies and produce statistical features comparable to experimental observations. 

\subsection*{Relationship to other network models}
Recurrent network models have been used in other contexts to study how networks of neurons perform specific tasks or simulate neural activity \cite{Renart2010, Toyoizumi2011, Laje2013, Mante2013, Sadeh2015, Morcos2016, Rajan2016}. By comparison with such models, our model is exceedingly simple: it is a sparsely connected, random network of excitatory and inhibitory neurons with a firing rate non-linearity. In this network, temporal information (about stimulus frequency) is transformed into a spatial representation. There are a number of ways to make this model more realistic. For instance, spatially distinct neural representations could trigger distinct neural trajectories \cite{Rajan2016}, matching the spatio-temporal multineuronal dynamics on single trials more closely. We did not pursue this here, as our goal was to show that heterogeneity in network connectivity could explain many features of population recordings during a simple discrimination task.

One of our key results is that selectivity can emerge for task parameters from an unstructured, random network. Several theoretical studies have previously examined the emergence of strong selectivity in random networks.  For example, it has long been recognized that orientation selectivity in primary visual cortex could emerge from random projections from geniculate inputs \cite{Malsburg1973, Ringach2004} and, in balanced state networks, this selectivity would be robust due to the dynamic cancellation of non-selective inputs \cite{Sadeh2015, Pattadkal2018, Hansel4049, Pehlevan2014}.  Moreover, selectivity from random projections could be enhanced through learning mechanisms either in the feed-forward projections \cite{Malsburg1973} or in the recurrent cortical network \cite{Sadeh2015}.  The mechanism for selectivity presented here adds to these by demonstrating selectivity for parameters that reflect temporal, rather than spatial, patterns. In this model, there is no spatial heterogeneity in the inputs, and the source of selectivity is not from heterogeneity in feed-forward projections, but from the dynamics of a recurrently connected non-linear network. 

Finally, our theory makes the following prediction, which should be verifiable experimentally. Suppose, the decision boundary for reporting low- versus high-frequency stimuli changed. If the network is structured as separate pools of excitatory and inhibitory neurons representing choice for one or the other stimulus category, then the representation of choice in PPC will not change with the task. If instead functional organization is generated by emergent network properties, when the task changes, the selectivity of individual cells will shift, as different pools of neurons represent the low- versus high-frequency stimuli. These pools of neurons would be overlapping, as the frequencies being discriminated change. 

As ever larger populations of neurons are simultaneously recorded, and experiments frequently focus on a variants of well-controlled sensory discrimination tasks, we face a tremendous challenge in inferring mechanism from observations. Very generally, there are two mechanisms that could account for diverse and mixed selectivity along with patterned functional connectivity across a population of neurons engaged in some experimental task. The first is that the network is wired specifically to achieve this, and if that is the case, then one must also explain the developmental or learning process that produced such intricate topology in the network. The alternative, which we explored here, is that the observed patterns of  selectivity can be explained as an emergent phenomena from simple patterns of statistical connectivity. In the particular case examined here, we were able to reproduce distributions of selectivity, functional connectivity, and population readout weight patterns that were experimentally observed. Even though we analyzed the emergence of selectivity in a specific experiment, we believe that similar conclusions could hold in other applications, in which a broad distribution of selectivity is observed. 

% This is for arXiv purposes only and would be removed from the submission
\subsection*{Acknowledgments}
We thank the Kavli Foundation for supporting the Kavli Brain Forum  seminar series, which brought Dr.~Anne Churchland to Emory and put our attention to this problem. We thank Dr.~Dieter Jaeger and Dr.~Gordon Berman for insightful comments about the manuscript. This work was supported by NIH Grants R01NS084844 (AS and IN), R01EB022872, and R01NS099375 (IN), and by NSF Grant BCS-1822677 (IN). During the early conception of this project, AS was supported by NIH/NINDS U01NS094302 and R01NS104928. 

\section*{Materials and methods}
\subsection*{Network simulation}
\subsubsection*{Random recurrent firing rate network}
We simulated a recurrent neural network of $N = 500$ neurons ($N_E = 400$ excitatory neurons and $N_I = 100$ inhibitory neurons) using standard firing rate equations:
\begin{align}\label{eqn:network}
    \tau \dot{\mathbf{x}} &= J\mathbf{r} + \mathbf{c}i(t) - \mathbf{x}, \\
    r_i &= g(x_i), \\
    g(x) &= 0.5*(1 + \tanh(x - b)),
\end{align}
where $x_i$ is the ``membrane potential'' of neuron $i$ and $r_i$ is its firing rate, obtained from $x_i$ through the nonlinear transfer function $g$. We set $\tau$ to 1 so all time is measured in units of the unit time constant. The transfer function $g$ ensures activity is between $0$ and $1$ for all cells. We included a bias term $b$, set to $2$, and this ensured small (0.05 or less) spontaneous ($i(t)=0$) firing rates in steady state. Neurons interacted through the matrix $J$, which was sparse ($20\%$ nonzero) and defined whether neurons are excitatory ($80\%$ of cells) or inhibitory. Weights originating from excitatory (inhibitory) neurons were drawn from a normal distribution with mean $g_E = 0.18$ ($g_I = -0.72$) and standard deviation $0.045$ (both excitatory and inhibitory), which balanced excitatory and inhibitory synaptic inputs on average across the population. Synaptic delays are not modeled. Finally, the external stimulus was the product of a scalar function $i(t)$ capturing the impulses (described below) and the binary input vector $\mathbf{c}$. The vector $\mathbf{c}$ was 1 for the excitatory cells that received direct inputs ($20\%$ of cells, randomly selected) and otherwise $0$, and this vector was fixed. In other words, the same subset of cells received inputs for all stimuli. 
The network was simulated using custom-written code in Matlab.  

\subsubsection*{The frequency discrimination task}
In the simulated task, the network was driven by stimuli consisting of irregular impulses with average frequency $f$ (Fig.~\ref{fig:f1_setup}A). The length of the sampling period in the simulated task was $50\tau$, corresponding to an effective $\tau$ of 20~ms. In the first set of simulations, either $f = 8$~Hz or $f = 16$~Hz.  In simulations in Fig.~\ref{fig:f4_freqs}, we analyzed $f=10$~Hz to $f=20$~Hz.  

For each trial with average frequency $f$, the input times ($t_k$) were selected randomly with uniform probability from the stimulus interval ($50 \tau$, or $1000 ms$) by drawing a fixed number of time points (e.g., $8$). Temporal precision of impulse timing was $0.01\tau$, so impulse times were drawn from the integers $1$ to $5000$ without replacement. For simulated trials with the same impulse frequency, the trial-by-trial variation in input times was the only source of variability across trials. 

We assumed that upstream sensory networks filtered the pulse inputs, so the overall input current to the network was described by
\begin{equation}\label{eqn:inputcurrent}
    i(t) = \alpha \sum_{t_k < t} \frac{(t-t_k)^2}{a^2} \exp\left(-\frac{t-t_k}{a}\right),
\end{equation}
where $a$ is a filtering timescale of pre-processing networks and the summation was taken over all $t_k<t$. We set $a = 0.5$ for all simulations (full width at half max of a single pulse is $1.7\tau$). 

\begin{figure}
\includegraphics[width=1.1\textwidth, trim ={1.5cm 4cm 0cm 0cm},clip=true]{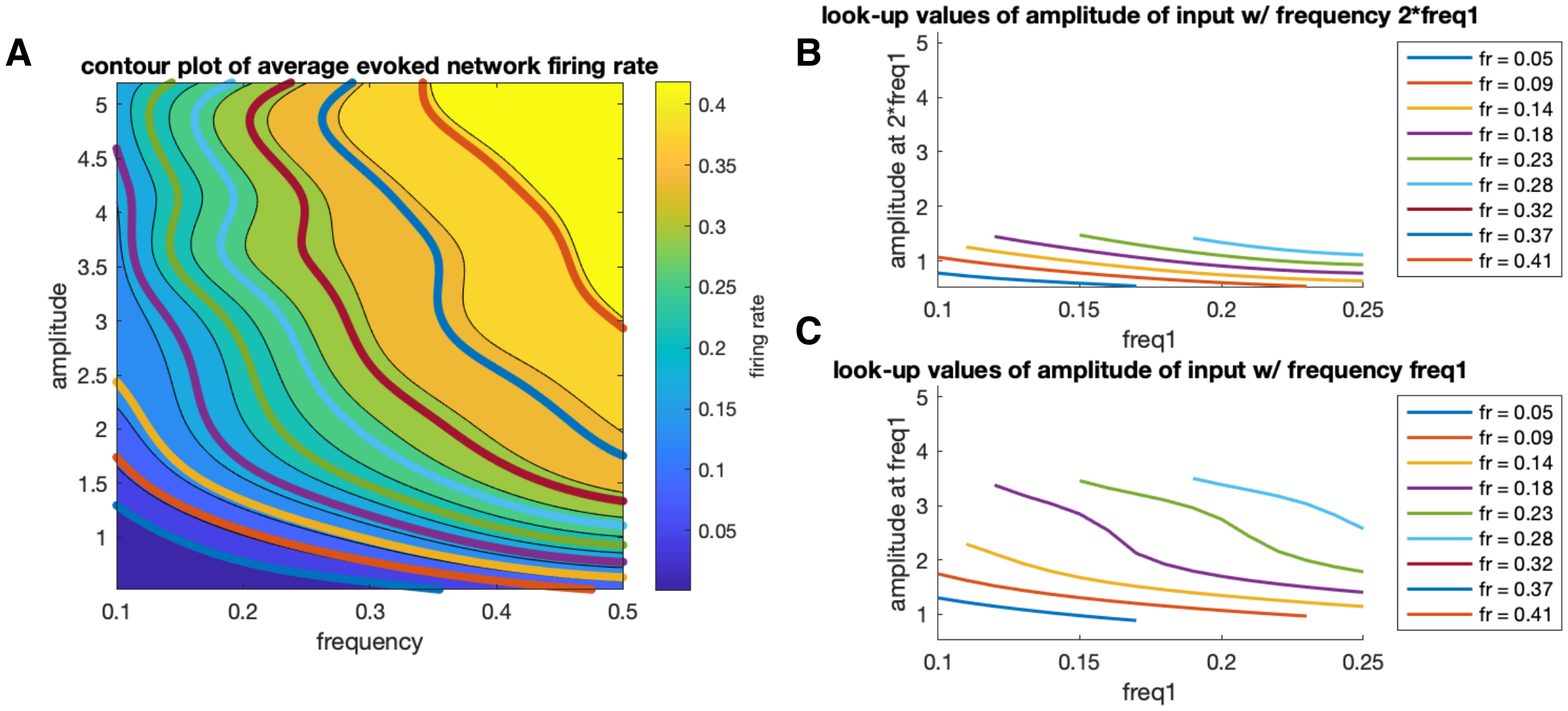}
\label{S1_Fig}

{\bf S1 Figure. Setting the average network firing rate.}  
A: Surface is a spline interpolation of average (over time and neurons) firing rate across 25 combinations of amplitude and frequency. Color on parula scale indicates firing rate (blue to yellow). Lines show contours at fixed average firing rate. 
For each firing rate (indicated by line color), we extract amplitudes and frequencies on the corresponding contour. 
B: Amplitudes for the high-frequency (freq2=2*freq1) input, plotted against the lower frequency (freq1). 
C: Amplitudes for the low-frequency input, plotted against the lower frequency (freq1). The highest firing rate contour that is defined over freq1 = 0.14 to 0.2 is simulated. Units of frequency are per $\tau$; multiplication by 50 converts to Hz. 
\end{figure}

Input currents were scaled by a frequency-dependent factor $\alpha$ to match the average firing rate across the network between conditions (Fig.~\ref{fig:f1_setup}, \nameref{S1_Fig}). To implement this scaling, 25 sets of input parameters (frequency and amplitude) were simulated with input frequencies from $0.1$ to $0.5$ (per $\tau$), and amplitudes from $0.3$ to $15$, a range spanning parameters that generated a range of firing rates in the network. From each of these 25 simulations, we computed the average network firing rate, and we interpolated this surface to find contours of equal firing rate. For the simulated experiment, we used combinations of frequency and amplitude that fall on a fixed contour. Across networks, the typical amplitude ratio between low- and high-frequency inputs was $2.1$. We verified post-simulation that the average firing rates match across frequency conditions (see, e.g., Fig.~\ref{fig:f1_setup}D).   

\subsection*{Simulation Analysis}
\subsubsection*{Single neuron selectivity}
For a pair of stimuli (e.g., 8 Hz and 16 Hz), we used an ideal observer to determine selectivity in single neurons. For each cell, the area under the receiver-operator curve (AUC) was computed nonparametrically from the distribution of low-frequency responses and the distribution of high-frequency responses at the end of the stimulus period \cite{Green1966}. Significance bounds are the 2.5-97.5 percentiles of the trial-shuffled distributions. Single neurons were selective if their AUC value fell outside the significance bounds. 

\subsubsection*{Noise correlations}
We computed noise correlations from the activity at the end of the stimulus period by subtracting the stimulus-averaged response and then computing neuron-neuron correlation coefficients. 

\subsubsection*{Classifier analysis}
The goal of the classification was to discriminate between two input frequencies using the simulated activity patterns. Simulated network activity was temporally down-sampled by averaging over time windows of size 1 ($\tau$). Neurons that received direct inputs were excluded from the decoding analysis, leaving 320 excitatory neurons and 100 inhibitory neurons. We fit classifiers separately on the full population (``all,'' 420 neurons), a subset of excitatory neurons (``exc-sub'', 100) and a subset of inhibitory neurons (``inh'', 100). We split the 800 simulated (400 in each condition) into ``test’’ and ``training’’ sets. We trained a classifier (linear kernel, SVM) to predict the stimulus label based on the activity (in ``all,’’ ``exc-sub’’ and ``inh’’) at each time point. We trained the classifier on the z-scored activity from each cell \cite{Licata2017}:
\begin{equation}
z_i = \frac{r_i - \bar{r}_i}{\sigma_{r_i}}.
\end{equation}
The classifier finds a rule $\mathbf{\xi}, \eta$ 
\begin{eqnarray}
\mathbf{\xi}*\mathbf{z} > \eta .
\end{eqnarray}
We also calculated the weights ($\mathbf{w}$) and bias ($b$) that operate on firing rates directly
\begin{eqnarray}
\mathbf{w}*\mathbf{r} > b.
\end{eqnarray}
Classifier accuracy was calculated on the reserved test set. In Fig.~\ref{fig:f3_decode}A, a classifier was fit at a single time point at the end of the stimulus window and tested at all other time points. Uncertainty in classifier accuracy was estimated by fitting the classifier using different cuts of the data: train/cross-val/test sets were drawn randomly, a classifier fit, and weights and accuracy on the test set recorded. This was repeated 50 times, and the error bar on classifier accuracy is the standard deviation across test set accuracy generated in this way.

% how to reference Appendix items: \nameref{S1_Appendix}.

\section*{Supporting information}

% % Include only the SI item label in the paragraph heading. Use the \nameref{label} command to cite SI items in the text.
\paragraph*{S1 Fig.}

\paragraph*{S2 Fig.}
\newpage

\bibliography{rrnei_refs.bib}

\end{document}